# Anomalous Raman scattering in layered AgCrP$_2$Se$_6$: Helical modes and excitation energy-dependent intensities

Rahul Rao,[1]* Jie Jiang,[1,2] Ruth Pachter,[1] Thuc T. Mai,[1,2] Valentine Mohaugen,[1] Maria F. Muñoz,[3] Ryan Siebenaller,[1,4] Emmanuel Rowe,[1] Ryan Selhorst,[1] Andrea N. Giordano,[1,5] Angela R. Hight Walker,[3] Michael A. Susner[1]

[1]Materials and Manufacturing Directorate, Air Force Research Laboratory, Wright-Patterson Air Force Base, OH 45433, USA

[2]Blue Halo Inc., Dayton OH 45434, USA

[3]Quantum Measurement Division, Physical Measurement Laboratory, NIST, Gaithersburg, MD 20899

[4]Department of Materials Science, The Ohio State University, Columbus, OH 43210, USA

[5]National Research Council, Washington, D.C., 20001

*Correspondence: Rahul.rao.2@us.af.mil

**Abstract**

Structural anisotropy in layered two-dimensional materials can lead to highly anisotropic optical absorption which, in turn, can profoundly affect their phonon modes. These effects include lattice orientation-dependent and excitation energy-dependent mode intensities that can enable new phononic and optoelectronic applications. Here, we report anomalous Raman spectra in single-crystalline AgCrP$_2$Se$_6$, a layered antiferromagnetic material. Density functional theory calculations and experimental measurements reveal several unique features in the Raman spectra of bulk and exfoliated AgCrP$_2$Se$_6$ crystals including three helical vibrational modes. These modes exhibit large Raman optical activities (circular intensity differences) in bulk AgCrP$_2$Se$_6$, which progressively decrease with thickness. We also observe strong excitation energy dependent peak intensities as well as a decrease in anti-Stokes peak intensities at room temperature with increasing excitation energy, resulting in an apparent cooling by up to 220 K. All of these anomalies in bulk and exfoliated flakes are attributed to the unique ABC layer stacking structure of $AgCrP_2Se_6$ and to the smaller unit cell volume that causes hybridization between the Se and Ag/Cr electron densities, resulting in charge transfer and strongly affecting the electron-phonon coupling. This work thus positions $AgCrP_2Se_6$ as an exciting new 2D material for optical and phononic applications.

**Introduction**

Structural anisotropy in layered two-dimensional (2D) materials can lead to highly anisotropic optical absorption, enabling new ways to tune light-matter coupling through the excitation, detection and control of light along various crystallographic axes.[1–6] Some of the consequential effects are hyperbolic plasmons and excitons,[7,8] giant non-linear second harmonic generation,[9] as well as mechanical anisotropy.[10] These properties can be leveraged in various photonic, phononic and optoelectronic devices wherein the 2D materials are incorporated as the active layer. Examples of applications using such devices include optical birefringent polarizers, polarized light emitting diodes and lasers, polarization sensitive photodetectors and thermoelectrics.[2,11]

The anisotropic optical response of 2D materials can be readily measured with angle-resolved linearly polarized Raman scattering; prior studies have revealed significant lattice orientation-dependent mode intensities in materials like black phosphorus,[12,13] $ReSe_2$,[14,15] $MoO_3$,[16] $As_2S_3$,[17] and $WTe_2$.[18] The Raman peak intensities can also depend on the energy/wavelength of the excitation laser and can be studied using resonance Raman scattering, where the incident photon absorption results in a real electronic transition rather than a virtual transition as in the normal Raman scattering process. Mode-specific excitation energy-dependent intensities have been found for Raman-active Stokes and anti-Stokes modes in $ReS_2$[19] and $MoTe_2$[20,21]. They have been attributed to larger contributions of electron–phonon coupling for specific phonon modes compared to electron–photon interaction. In addition, circularly polarized Raman scattering studies have revealed large differences in the intensities of certain phonon modes in $ReS_2$ and $ReSe_2$ to either left or right circularly polarized (LCP and RCP, respectively) excitation.[22] These differences were attributed to quantum interference between the first-order Raman scattering processes at different *k*-points in the Brillouin zone.[23] In short, optical anisotropies may affect the vibrational modes of 2D materials in different ways, allowing for tunability of the electron-phonon coupling in individual materials and their heterostructures.

Here, we report anomalous Raman spectra in single-crystalline AgCrP$_2$Se$_6$, which is a layered antiferromagnetic material (T$_N$ ~42K)[24] belonging to the 2D metal phosphorus chalcogenide/metal thio- and seleno-phosphate family.[25] Theoretical calculations and experimental measurements revealed several unique features in the Raman spectra of bulk and exfoliated AgCrP$_2$Se$_6$ crystals including three helical vibrational modes. These modes exhibit large Raman optical activities (circular intensity differences) in bulk AgCrP$_2$Se$_6$, which progressively decrease with thickness. We also observed variations in peak intensities in the bulk crystal and exfoliated flakes, some of which are maximum at our lowest excitation energy (1.58 eV) and others maximized at higher excitation energies (2.33 or 2.41 eV). Finally, we observed a decrease in anti-Stokes peak intensities at room temperature with increasing excitation energy, resulting in an apparent cooling by up to 220 K. We attribute our observations to the structural anisotropy in AgCrP$_2$Se$_6$. The helical modes arise from the unique ABC layer stacking and the anomalous peak intensities are tentatively attributed to the smaller unit cell volume that causes hybridization between the Se and Ag/Cr electron densities and consequently charge transfer and variations in electron-phonon coupling.

**Results and Discussion**

**Observation of helical Raman modes**

Each layer of AgCrP$_2$Se$_6$ consists of a [P$_2$Se$_6$]$^{4-}$ framework in which the Se atoms are positioned on the vertices of Se$_6$ octahedra. One third of these octahedral sites contain P-P dimers which support the anionic sublattice; the remaining two thirds of the octahedral sites are occupied by hexagonally arranged and alternating Ag$^+$ and Cr$^{3+}$ cations. Our previous study[24] showed the crystal structure of AgCrP$_2$Se$_6$ to be trigonal, belonging to the *P*3$_1$12 space group (No. 151). The schematic in Figure 1a shows four layers (one unit cell plus an additional layer), each with a [P$_2$Se$_6$]$^{4-}$ unit and Ag$^+$ and Cr$^{3+}$ ions. Note that *P*3$_1$12 is a chiral space group and that inorganic materials in this space group can exhibit structural chirality[26] as well as chiral phonons.[27] Indeed, a helical arrangement of the Ag$^+$ and Cr$^{3+}$ ions can be seen in the structure along the layer stacking direction (denoted by the arrows in Figure 1a). Further confirmation of the helical atomic arrangement can be seen in a top-down view of the Ag$^+$ and Cr$^{3+}$ ions as well as a [P$_2$Se$_6$]$^{4-}$ unit, which shows C$_3$ rotational symmetry (Figure S1). The alternating hexagonal arrangement of metal cations is similar to other Ag- and Cu-based metal selenophosphates. However, a key difference between AgCrP$_2$Se$_6$ and related materials is the stacking sequence between the layers, which is ABC as can be seen in Figure 1a. The ABC stacking sequence can also be found in quaternary $CuAlP_2Se_6$[28] and ternary $In_{4/3}P_2S_6$,[29] $Cd_2P_2Se_6$, $Fe_2P_2Se_6$, $Mg_2P_2Se_6$, $Mn_2P_2Se_6$, and $Zn_2P_2Se_6$;[30] related 6 or 12 layer stacking sequences are also observed in the lower temperature phases of $CuBiP_2Se_6$.[31] This ABC stacking is responsible for the helical arrangement of the atoms along the stacking direction, and as discussed further below, has an important influence on the vibrational modes of $AgCrP_2Se_6$.

Figure 1b shows unpolarized room temperature Raman spectra from bulk $AgCrP_2Se_6$, collected with multiple laser excitation energies ($E_{laser}$ = 1.58, 1.96, 2.33, 2.41 and 2.54 eV), along with the spectrum calculated using density functional theory (DFT, bottom spectrum in Figure 1b). In all, 11 Raman modes can be observed in the Raman spectra and are labeled $P_1$-$P_{11}$. The peaks are denoted in Figure 1b by the solid vertical lines underneath the spectrum collected with 1.58 eV excitation. All the 11 peaks are captured and labeled in the calculated spectrum. The slight discrepancies between the calculated and experimental peak frequencies can be attributed to the level of theory applied (PBE+D3 functional, more details in the Experimental Methods). The 11 Raman modes correspond to either A or E symmetry, which were confirmed by collecting linearly polarized Raman spectra (Figure S2).

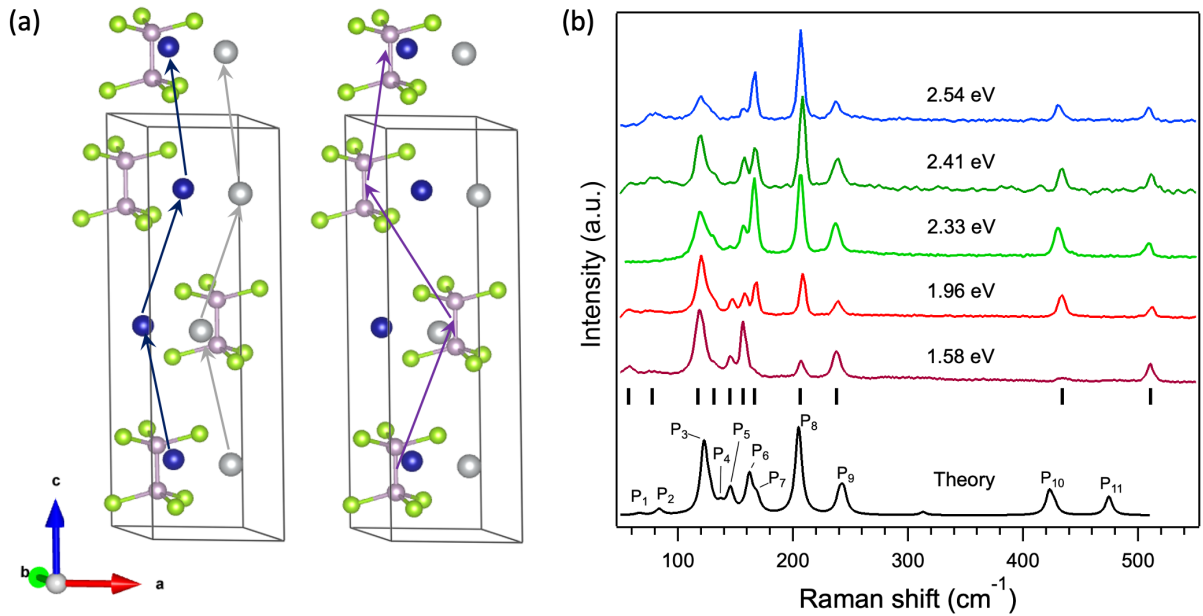

Figure 1. a) Schematic showing 4 layers of $AgCrP_2Se_6$. The arrows indicate right-handed helical arrangement of the $Ag^+$ and $Cr^{3+}$ ions, and $P_2Se_6$ atom groups along the *c*-axis. The Ag, Cr, P, and Se atoms are depicted in grey, blue, purple and green, respectively. b) Experimentally measured multi-excitation (1.58 – 2.54 eV) Raman spectra collected at room temperature from bulk $AgCrP_2Se_6$ crystals along with the calculated spectrum (bottom trace).

The phonon eigenvectors for the eleven Raman modes are shown in Figure 2. Of the eleven modes, three are helical ($P_1$, $P_5$ and $P_9$, all modes with E symmetry). $P_1$ corresponds to helical vibrations of the $Ag^+$ ions, and $P_5$ and $P_9$ correspond to helical vibrations of the $Cr^{3+}$ ions. $P_5$ also involves in-plane vibrations of the Se atoms. For the remaining non-helical modes, $P_3$ corresponds to in-plane shear-like vibrations of the Se atoms, $P_8$ is a Se in-plane breathing mode, and $P_6$ is an out-of-plane breathing mode of the Se atoms along the *c*-axis and perpendicular to the layer stacking direction. The peaks denoted as $P_2$, $P_4$, and $P_7$ correspond to Se atom vibrations with both in-plane and out-of-plane components. The highest frequency modes $P_{10}$ and $P_{11}$ correspond to in-plane and out-of-plane vibrations of the P atoms, respectively. For clarity, the eleven Raman-active modes measured experimentally along with the DFT-calculated mode frequencies and their corresponding symmetries are listed in Table 1.

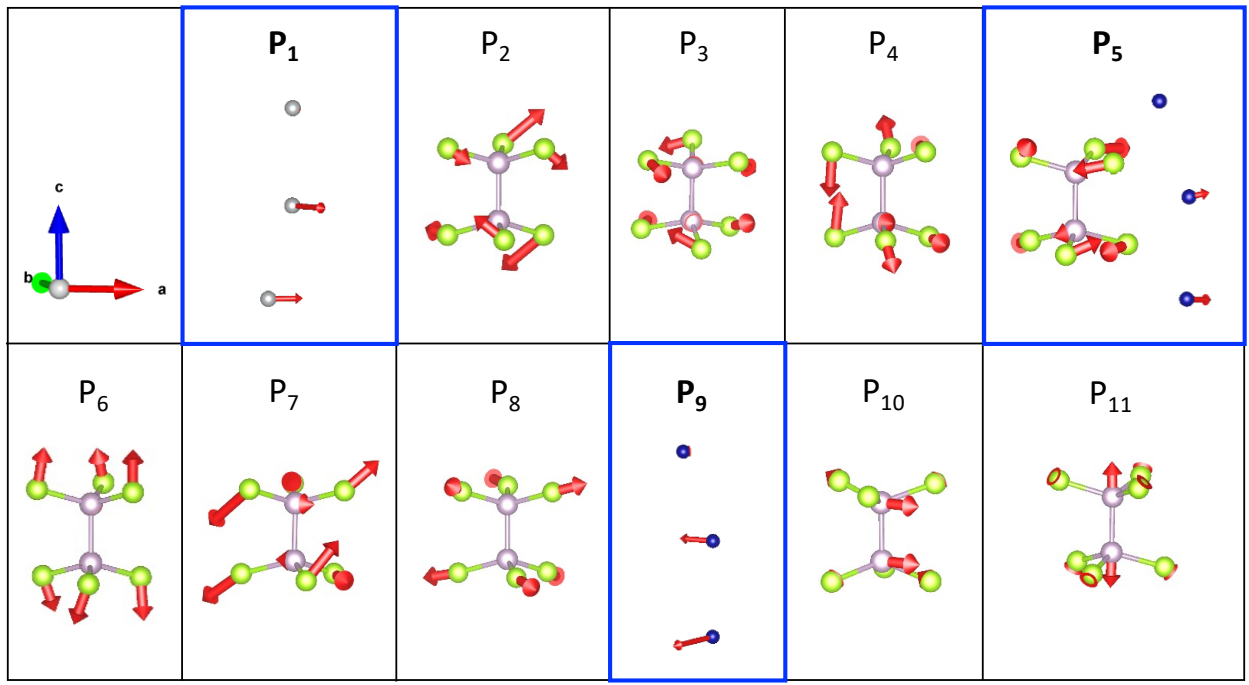

Figure 2. Phonon eigenvectors for the 11 observed Raman peaks ($P_1$-$P_{11}$) in $AgCrP_2Se_6$. The Ag, Cr, P and Se atoms are in silver, blue, violet and green, respectively. The three helical modes ($P_1$, $P_5$ and $P_9$) are highlighted by the blue boxes.

The presence of helical vibrations in $AgCrP_2Se_6$ is unique and we confirmed their helicity by measuring circularly polarized Raman spectra. Figure 3a shows circularly polarized Raman spectra collected from a bulk AgCrP$_2$Se$_6$ crystal using $E_{laser}$=1.58 eV (785 nm). Left circular polarized (LCP, σ$^+$) incident light was achieved by manually inserting a half waveplate in combination with a quarter waveplate prior to the objective lens in our micro-Raman setup (the optical layout of the circular polarization measurement setup is shown in Figure S3), while the half waveplate was moved out of the beam path to achieve right circularly polarized (RCP, σ$^-$) excitation. In our experimental setup, the backscattered light is directed through the notch filter and is polarized parallel or perpendicular, thus giving us co- and cross-circularly polarized configurations (σ$^+$σ$^+$ or σ$^-$σ$^+$, respectively). Figure 3a shows a clear difference in intensities of P$_1$, P$_5$ and P$_9$ between RCP and LCP excitations. While the other peaks do not vary in intensity with RCP or LCP excitation, the intensities of P$_1$, P$_5$ and P$_9$ are higher in the RCP spectrum. The variations in intensity due to differences in absorption of RCP or LCP light confirms their helical nature. The notch filter also allows us to measure both Stokes and anti-Stokes peaks; we observe similar differences in intensities between the RCP and LCP spectra in the anti-Stokes region (Figure S4).

Table 1. Comparison between measured and calculated Raman peak frequencies (in cm$^{-1}$) in AgCrP$_2$Se$_6$ along with their corresponding mode symmetries. The three helical modes P$_1$, P$_5$ and P$_9$ are indicated in bold font.

| Raman peak | **P$_1$** | P$_2$ | P$_3$ | P$_4$ | **P$_5$** | P$_6$ | P$_7$ | P$_8$ | **P$_9$** | P$_{10}$ | P$_{11}$ |
|---|---|---|---|---|---|---|---|---|---|---|---|
| Experiment | **65** | 83.5 | 120 | 129 | **147.5** | 158 | 166.3 | 208.1 | **239.3** | 434.4 | 511.1 |
| Theory | **65.42** | 83.60 | 122.30 | 137.75 | **145.66** | 162.00 | 168.49 | 204.81 | **243.97** | 423.04 | 474.51 |
| Mode symmetry | **E** | E | A | A | **E** | A | E | A | **E** | E | A |

Chiral phonons (i.e. circular motions of atoms perpendicular to the direction of propagation of the phonon mode) have been observed previously in chiral crystals such as α-quartz,[32] tellurium[33] and HgS.[34] The chiral phonons in these materials are degenerate E modes with opposite phonon pseudo-angular momenta (PAM, $l_{ph}= \pm1$) unlike the A phonon modes, which have one dimensional irreducible representation and with $l_{ph}=0$.[35] The condition of $l_{ph}= \pm1$ results in a splitting of the degenerate E modes (typically by a few cm$^{-1}$) at low wavevectors away from the Brillouin zone center, enabling their measurement with circularly polarized Raman scattering. The helical E Raman modes ($P_1$, $P_5$ and $P_9$) in $AgCrP_2Se_6$ could demonstrate phonon chirality and future calculations considering the PAM could reveal whether or not these phonon modes are truly chiral. However, our calculated splitting between the modes is less than 0.1 cm$^{-1}$, well under the spectral resolution of our instruments. Figure S5 shows a higher magnification view of the phonon dispersion, showing the small degree of splitting between the degenerate E modes. It is worth noting that we do not see any differences in peak intensities with LCP and RCP excitation in the spectra from related materials $AgInP_2Se_6$ and $AgInP_2S_6$ (Figure S6). Both $AgInP_2Se_6$ and $AgInP_2S_6$ exhibit AB layer stacking,[36] suggesting that the structural anisotropy caused by the ABC layer stacking in $AgCrP_2Se_6$ is uniquely responsible for the presence of helical vibrational modes.

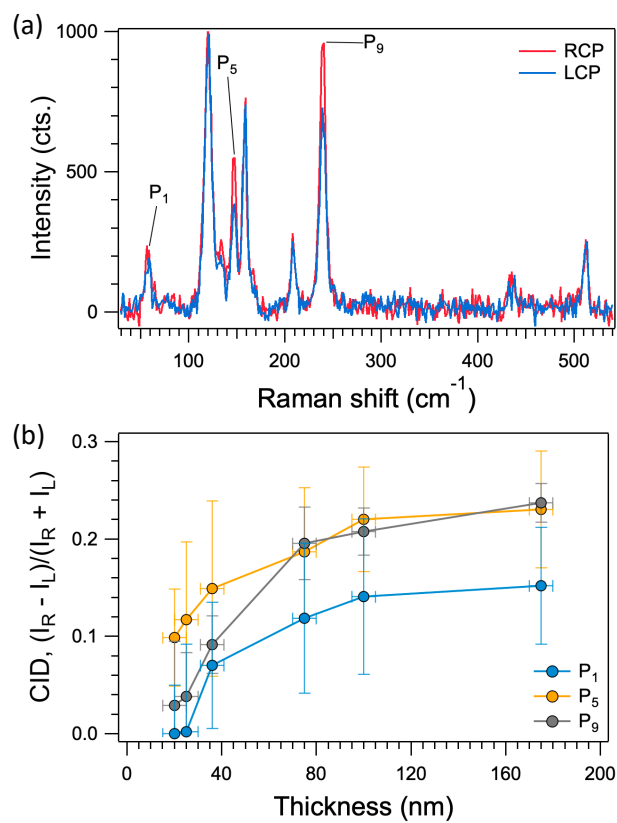

Figure 3. a) Circularly polarized Raman spectra from bulk $AgCrP_2Se_6$, collected with 785 nm (1.58 eV) excitation. b) Circular intensity difference as a function of flake thickness for the three helical modes $P_1$, $P_5$ and $P_9$.

The differences in spectral intensities with RCP and LCP excitation prompted us to measure the Raman optical activity (ROA) for the three helical vibrational modes. The ROA is defined as the difference in Raman scattering for RCP and LCP excitations, and its strength is estimated in the form of the circular intensity difference, $CID = (I_R - I_L)/(I_R + I_L)$.[37,38] $I_R$ and $I_L$ are the peak intensities corresponding to RCP and LCP excitation, respectively. Note that, in order to estimate $I_R$ and $I_L$, the circularly polarized Raman spectra were collected by placing a freshly cleaved $AgCrP_2Se_6$ crystal onto a Si substrate and the two spectra were normalized by considering the response of the Si substrate to the RCP and LCP excitation (under the same measurement conditions). In addition, we accounted for the loss in intensity of the LCP light due to the additional half waveplate in the excitation beam path. The CIDs for the three helical modes $P_1$, $P_5$ and $P_9$ (obtained from the Stokes Raman peaks for the spectrum plotted in Figure 3a) are 0.15, 0.21 and 0.20, respectively. These CID values are very high compared to what is typically observed for chiral materials, *i.e.* on the order of $10^{-3}$.[37] Recently, circularly polarized Raman studies on 2D $ReS_2$ and $ReSe_2$[22,23] reported similarly large CID values that were attributed to quantum interference effects. Note that in $ReS_2$ and $ReSe_2$, the phonon modes are not helical, but include a combination of in-plane and out-of-plane vibrations of the Re and chalcogen atoms. Thus, our measured CID values for intrinsically helical vibrational modes in $AgCrP_2Se_6$ are ostensibly the highest for a 2D material.

Next, in order to investigate the thickness-dependence of the CIDs in $AgCrP_2Se_6$, we mechanically exfoliated flakes down to a minimum thickness of 20 nm onto $Si/SiO_2$ (285 nm of oxide) substrates, followed by collection of circularly polarized Raman spectra. The thickness of the exfoliated flakes was confirmed using atomic force microscopy. For each thickness, the RCP and LCP spectra were normalized in the same manner as that described above for the bulk crystal. Figure 3b shows the CIDs for $P_1$, $P_5$ and $P_9$ as a function of thickness. Even with the sizable error bars, all three CIDs can be observed to decrease with thickness. This decrease can be tentatively attributed to a reduction in interlayer coupling with decreasing flake thickness. The trends also support the notion that the ABC stacking is crucial to the occurrence of the helical vibrational modes, and that the disruption of the ABC stacking with decreasing thickness results in the lower CIDs. Raman spectra from even thinner flakes were difficult to obtain owing to the drastically reduced intensities and increasing laser-induced degradation. Nonetheless, collectively the appearance of helical modes and their thickness dependence hint at the potential to manipulate them through moiré engineering or ion intercalation.

**Excitation energy-dependent mode intensities**

Going back to Figure 1b, it is clear that the intensities of several Raman peaks depend strongly on $E_{laser}$, and that this dependence varies for each mode. Interestingly, the non-helical Raman modes exhibit the greatest dependence on the excitation energy, with some peaks disappearing completely for certain values of $E_{laser}$. This can be seen more clearly in Figure 4, which plots the normalized intensities of the four non-helical modes that display the greatest variations in intensity (the excitation energy-dependent intensities of 9 of the 11 peaks are shown in Figure S7). Note that the multi-excitation spectra plotted in Figure 1b were collected by placing the crystal on a Si substrate and normalized by accounting for the wavelength-dependent absorption of the Si. Additionally, we were unable to measure the excitation intensity dependence of the lowest frequency modes ($P_1$ and $P_2$) due to instrumental limitations (lack of low-frequency cut-off filters for all the excitations).

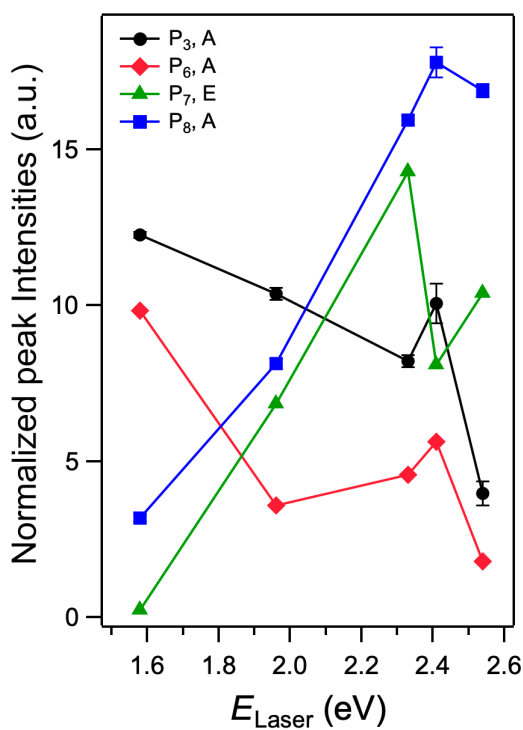

Figure 4. Normalized peak intensities for $P_3$, $P_6$, $P_7$ and $P_8$ as a function of excitation laser energy.

Three features can be gleaned from Figures 4 and S7. Firstly, the normalized intensities of $P_7$ and $P_8$ increase considerably with energy and are maximum at $E_{laser}$ = 2.41 and 2.33 eV, respectively; the intensity of $P_8$ is over a factor of 5 at $E_{laser}$ = 2.41 eV compared to its intensity at 1.58 eV. Secondly, the intensities of $P_3$ and $P_6$ are the highest for our lowest excitation energy (1.58 eV) and decrease with increasing $E_{laser}$. The third feature observed in Figures 4 and S7 is that the strong excitation energy dependence occurs for both E and A symmetry modes. Curiously, all four peaks that exhibit a strong excitation energy dependence ($P_3$, $P_7$, $P_6$ and $P_8$) correspond mainly to vibrations of the Se atoms. Our previous calculations[24] demonstrated that the Se *p* orbitals have large contributions to the valence and conduction bands. This could account for our observation of a larger dependence of the intensity on the excitation energy for the Se vibrational modes than for the higher frequency modes ($P_{10}$ and $P_{11}$, c.f. Figure S7) that mainly involve vibrations of the P atoms.

The dependence of Raman peak intensities on $E_{laser}$ typically arises because of resonance with an electronic state, where, instead of a virtual transition of the absorbed photon, there is a real electronic transition. Such increases in intensities of some Raman peaks at certain laser energies corresponding to excitonic transitions have been reported in $MoSe_2$[39] and $MoTe_2$.[20] In order to investigate this more closely, we calculated the optical absorbance spectrum for $AgCrP_2Se_6$, which is plotted in Figure 5a. Additionally, we measured the reflectance spectrum from a bulk crystal, as shown in Figure 5b. The dashed vertical lines in Figure 5 indicate all the laser excitation energies used in this study. The calculated absorbance spectrum in Figure 5a has peaks close to our laser excitation energies. However, these peaks are not sharp or intense. A similar lack of features can be seen in the experimental reflectance spectrum (Figure 5b), with a relatively flat response (27 - 28%) across the range of our laser energies. The low reflectance also indicates a high absorption of light across the entire visible wavelength range. We also observed the same dependence of the peak intensities on $E_{laser}$ for mechanically exfoliated flakes. The plots in figure S8 show Raman spectra collected with $E_{laser}$ =1.58, 1.96 and 2.41 eV where the excitation energy dependence of the peak intensities does not depend on the thickness. It is well known that the electronic structure of 2D materials depends strongly on their thickness and that, in general, a reduction in thickness is accompanied by an increase in the bandgap of the material.[40] The lack of sharp resonances at our laser energies as well as the independence of the peak intensities on flake thickness together suggest that resonance Raman enhancement is unlikely to be the primary reason for the our excitation energy-dependent observations (Figure 4).

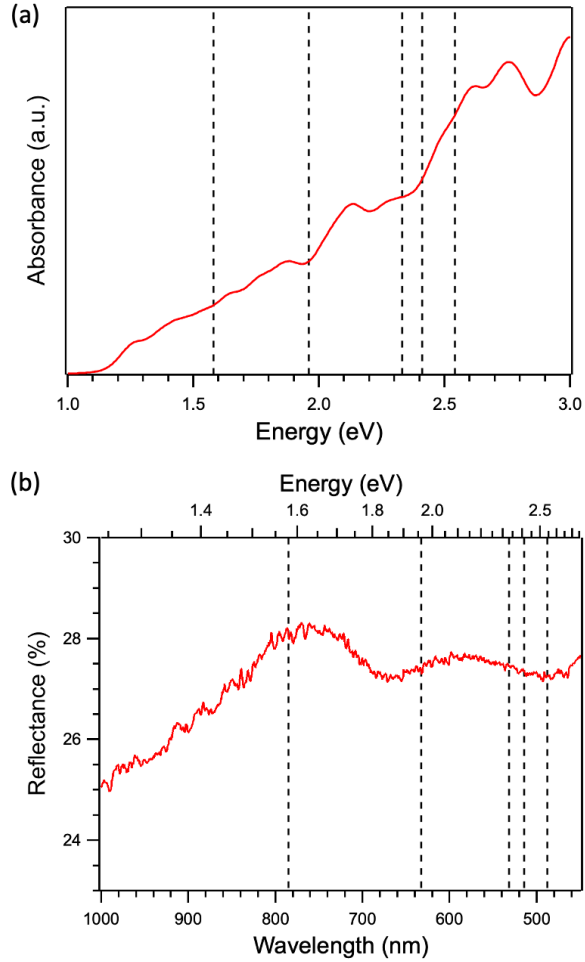

Figure 5. (a) Calculated absorption and (b) measured reflectance curves from $AgCrP_2Se_6$. The vertical dashed lines indicate all the laser excitation energies used in this study.

An alternate reason for mode-selective excitation energy-dependent Raman intensities could be the strength of the electron phonon coupling. Such enhancements have been reported before in other 2D materials like $ReS_2$[19] and $PdSe_2$,[41] where the structural anisotropy results in strong excitation energy dependences for peak intensities and cannot be accounted for by resonance with the excitation laser. We have previously shown that, among the 2D metal selenophosphates that have been synthesized thus far, $AgCrP_2Se_6$ possesses the lowest volume per $P_2Se_6$ unit (222 Å$^3$, c.f. Figure 1 in Ref. 24). The lower volume could lead to a larger overlap between the electron densities of the Se and the metal cations, thus enhancing charge transfer and electron-phonon coupling. We have previously observed such structural-induced variations in electron/spin-phonon coupling in $MPS_3$ materials (M = Ni, Co, Fe and Mn), where the lowest unit cell volume of NiPS$_3$ leads to a strong interaction between the S atoms and Ni$^{2+}$ ions, and thus a higher degree of electron-phonon coupling.[42] Indeed, a closer look at the calculated projected density of states (PDOS, Fig. S9) of AgCrP$_2$Se$_6$ reveals that the Ag 4$d$ bands hybridize with the Se 4$p$ bands close to the valence band maximum, while the Cr 3$d$ bands hybridize with the Se 4$p$ bands close to the conduction band minimum. The PDOS analyses also indicate a value of 3.97e for Cr$^{3+}$, which assumes a formal 3$d$ occupation number of 3 as per the electronic configuration of Cr. This is qualitatively consistent with self-doping due to charge transfer between the Se and Cr$^{3+}$ ions, similar to what has been reported previously for NiPS$_3$[43] (see Fig. S9 and the accompanying discussion). Our analysis thus hints at the strong influence of electron-phonon coupling on the Raman peak intensities. However, further multi-excitation spectral measurements using tunable laser sources as well as calculations of excitation energy-dependent spectra that include electron-phonon coupling effects may shed more light on our observations.

**Unusual anti-Stokes peak intensities**

Lastly, we report unusual trends in the anti-Stokes intensities in AgCrP$_2$Se$_6$. In the anti-Stokes scattering process, incident light absorption into the material results in the absorption/annihilation of phonons rather than the creation of phonons like in the Stokes process. As a result, the anti-Stokes Raman scattered light gains energy compared to the laser excitation energy. The phonon absorption occurs from excited state phonon populations, and thus the anti-Stokes process is closely correlated with temperature, with the anti-Stokes peak intensity increasing with temperature. The ratio of intensities between the anti-Stokes and Stokes Raman peaks ($I_{AS}/I_S$) is a commonly used metric that is related to the temperature (T) according to the following relation: $\frac{I_{AS}}{I_S} = \left(\frac{\omega_L+\omega_i}{\omega_L-\omega_i}\right)^4 \exp\left(-\frac{\hbar\omega_i}{kT}\right)$,[44] where $\omega_L$ and $\omega_i$ are the laser frequency and the frequency of the $i^{th}$ mode, respectively, $\hbar$ is the reduced Planck's constant and $k$ is the Boltzmann constant.

Figure 6a shows the Stokes and anti-Stokes regions (positive and negative frequencies, respectively) of unpolarized Raman spectra from bulk AgCrP$_2$Se$_6$ crystals, collected with $E_{laser}$ = 1.58, 1.96 and 2.41 eV. The spectra in Figure 6a were collected at room temperature with the lowest laser powers that produced measurable signals and reasonable signal/noise ratios (64, 105 and 68 μW for $E_{laser}$ = 1.58, 1.96 and 2.41 eV, respectively). We fit the Raman peaks in the spectra in Figure 6a with Voigt lineshapes to obtain the anti-Stokes and Stokes intensities ($I_{AS}$ and $I_S$, respectively), whose ratios are plotted in Figure 6b against $E_{laser}$ for four Raman peaks – $P_3$, $P_6$, $P_8$ and $P_9$. The figure shows that the measured $I_{AS}/I_S$ ratios are highest for $E_{laser}$ = 1.58 eV and decrease with increasing $E_{laser}$ for all four peaks.

Excitation energy-dependent variations in $I_{AS}/I_S$ ratios can arise from resonance effects, where the excitation laser is in resonance with either the Stokes or anti-Stokes photons. In other words, the resonance occurs for energies that are greater or lower than the laser energy by the phonon energy. In our case the highest phonon mode in Figure 6a ($P_9$ ~240 cm$^{-1}$) corresponds to 29.7 meV higher or lower than the laser energy. As shown in Figures 5a and 5b, we do not see any sharp features in the optical spectrum of $AgCrP_2Se_6$ between 500 – 800 nm that could account for resonances with the Stokes or anti-Stokes scattered photon energies. Moreover, $AgCrP_2Se_6$ is an indirect gap semiconductor (bandgap ~1.24 eV)[24] and we do not expect resonance at the bandgap energy. Thus, even though our lowest excitation energy (1.58 eV) is considerably higher than the bandgap, the trend of increasing $I_{AS}/I_S$ ratios with decreasing laser energy cannot be attributed to laser excitation with energies close(r) to resonance. The unusual laser energy dependence of the $I_{AS}/I_S$ ratios are likely attributable to a different origin.

Also plotted in Figure 6b are the theoretically expected (according to the above relationship) $I_{AS}/I_S$ ratios at 298 K. The dashed line in Figure 6b is the average calculated $I_{AS}/I_S$ ratio and the shaded region represents the spread in values for the various laser energies. While there are slight variations in the laser energy dependence of the $I_{AS}/I_S$ ratios between the peaks, on the whole the $I_{AS}/I_S$ ratios are slightly above the expected room temperature values for $E_{laser}$ = 1.58 eV and lower than expected for $E_{laser}$ = 1.96 and 2.41 eV. The lower $I_{AS}/I_S$ ratios signify an apparent cooling of the phonon modes. Again, using the above relationship, we calculated the temperatures from the $I_{AS}/I_S$ ratios. Apart from the highest frequency $P_9$ mode ~240 cm$^{-1}$, the temperatures inferred from the $I_{AS}/I_S$ ratios for the other peaks ($P_3$, $P_6$ and $P_8$) range from 83 to 256 K, all significantly below room temperature. The inferred temperatures obtained from the $I_{AS}/I_S$ ratios for the four Raman modes are plotted in Figure 6c. We also measured the $I_{AS}/I_S$ ratios in mechanically exfoliated flakes as a function of thickness, and they are plotted in Figure 6d. On the whole, the $I_{AS}/I_S$ ratios decrease with thickness for both of the measured laser excitations (1.58 and 1.96 eV).

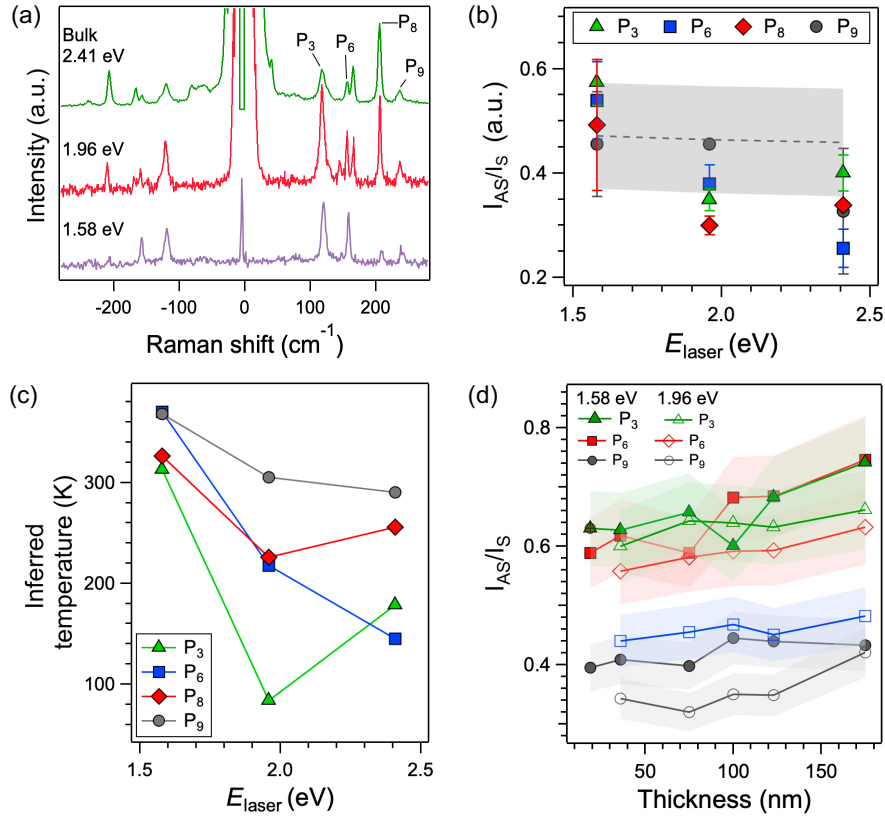

Figure 6. a) Room temperature multi-excitation Raman spectra in the Stokes and anti-Stokes region from AgCrP$_2$Se$_6$ crystals. b) $I_{AS}/I_S$ ratios for four Raman modes (P$_3$, P$_6$, P$_8$, P$_9$) as a function of $E_{laser}$. The dashed line is the calculated $I_{AS}/I_S$ ratios corresponding to room temperature (298 K), and the grey shaded area represents the spread in values for the three laser excitation energies. c) Temperatures calculated from the measured $I_{AS}/I_S$ ratios, plotted against $E_{laser}$. d) $I_{AS}/I_S$ ratios as a function of AgCrP$_2$Se$_6$ flake thickness. The filled and open data points correspond to $E_{laser}$ = 1.58 and 1.96 eV, respectively. The errors in estimation of the $I_{AS}/I_S$ ratios are plotted as error surfaces.

The phenomenon of laser-induced cooling by the removal of phonons is well-known and has been mainly observed in ultrafast Raman spectroscopy measurements.[45,46] However, our measurements were performed under steady state conditions, wherein the energy gained by removing phonons must be somehow dissipated. Laser-induced cooling in steady state (under continuous wave excitation) was previously reported for individual semiconducting single-walled carbon nanotubes where the low $I_{AS}/I_S$ ratios were attributed to a phonon-assisted inter-band relaxation and luminescence emission process.[47] However, we do not observe any luminescence at energies other than across the band edge in AgCrP$_2$Se$_6$,[24] suggesting that this mechanism is not responsible for our observed low $I_{AS}/I_S$ ratios for $E_{laser}$ = 1.96 and 2.41 eV. To gain more insights, we performed laser power- and temperature-dependent Raman measurements. The $I_{AS}/I_S$ ratios exhibit an exponential dependence on laser powers (Figure S10), following the Boltzmann phonon population distribution equation and implying that the anti-Stokes phonon populations are thermally driven with increasing laser powers. Interestingly, for the highest laser power (1.78 mW) with the 1.58 eV excitation, the calculated temperature from the $I_{AS}/I_S$ ratios is over 1100 K (Figure S11), which is above the typical decomposition temperature of this class of materials (1000 - 1100 K).[29,48] We confirmed the structural stability of the material by collecting a Raman spectrum after lowering the power.

To address the origin of the unusual Stokes and anti-Stokes intensities in $AgCrP_2Se_6$ and its excitation energy dependence, we first note that the calculated absorption spectrum and the experimental reflectance spectrum both show increasing light absorption at shorter wavelengths. Secondly, we turn back to the discussion in the previous section, namely, the hybridization between the Se and Cr orbitals resulting in charge transfer between the Se and $Cr^{3+}$ ions. We propose that the greater light absorption at higher laser excitation energies heightens this charge transfer, and the depopulation/annihilation of the vibrational modes compensates for the energy needed for this charge transfer. Such a phenomenon was recently observed in steady state surface-enhanced Raman spectroscopy measurements from aromatic thiols adsorbed on plasmonic gold nanoparticles,[49] wherein a Marcus charge transfer process was postulated as the reason for the apparent cooling of the molecule by ~60 K. Lastly, the decreasing $I_{AS}/I_S$ ratios with thickness (Figure 6d) mirror that of the CIDs (Figure 3b) and could be attributed to a reduction in the interlayer ABC coupling, causing a reduction in this charge transfer. Further ultrafast spectroscopic measurements as well as excitation energy-dependent computations may help to explain our observations.

**Conclusions**

We have computed and measured the Raman spectrum from bulk and exfoliated $AgCrP_2Se_6$, and observe several unusual features – 1. Three of the eleven Raman-active modes are helical, and were confirmed with circularly polarized Raman measurements. These modes exhibit large circular intensity differences that decrease with thickness in the exfoliated flakes, 2. Several modes exhibit distinct excitation energy-dependent intensities, and 3. We observe low anti-Stokes Raman peak intensities with increasing laser excitation energy, which results in an apparent cooling of the material by up to 220 K. We attribute these anomalies to the unique ABC layer stacking structure of AgCrP$_2$Se$_6$ and to the smaller unit cell volume that causes hybridization between the Se and Ag/Cr electron densities, resulting in charge transfer, and strongly affecting the electron-phonon coupling. Coupled with our previous observation of sharp defect-induced luminescence ~696 nm,[24] this work establishes AgCrP$_2$Se$_6$ as an exciting new 2D material for a variety of optical and phononics applications.

**Acknowledgements**

We acknowledge funding support from the Air Force Office of Scientific Research (AFOSR) grant no. LRIR 23RXCOR003, AOARD MOST grant no. F4GGA21207H002 and from the National Research Council's Senior NRC Associateship program. We also thank Prof. Riichiro Saito and Prof. Reneé Frontiera for illuminating discussions.

**Experimental Section**

AgCrP$_2$Se$_6$ single crystals were synthesized by combining Ag foil (Johnson Matthey, 99.99%), Cr powder (Aldrich, >99.5%), Se shot (Alfa Aesar Puratronic, 99.999%), and P lumps (Alfa Aesar Puratronic, 99.999%) together with ~100 mg of I$_2$ (Alfa Aesar ACS, 99.8%) in a sealed quartz ampoule. The ampoule was inserted into a quartz tube furnace which was heated to a temperature of 650 °C and held for 200 h to drive the reaction to completion. The crystals were micaceous and presented as well-faceted distorted hexagons mostly 20–200 μm in diameter, though a few grew to 2 mm and were reserved for magnetic and optical characterization; thicknesses were ~20–40 μm. The stoichiometry and crystal structure of the product was confirmed by Electron Dispersive Spectroscopy (Thermo Scientific Ultra Dry EDS spectrometer joined with a JEOL JSM-6060 scanning electron microscope), which revealed the composition to be Ag$_{1.13(8)}$Cr$_{0.94(10)}$P$_{1.87(9)}$Se$_{6.07(10)}$, within error to the ideal composition of AgCrP$_2$Se$_6$.

Density functional theory (DFT) calculations were performed with the Vienna *ab initio* simulation package (VASP 5.4), applying the projector augmented-wave potential (PAW). The Kohn-Sham equations were solved using a plane wave basis set with an energy cutoff of 500 eV. We employed our experimentally determined structure as the initial structure and optimized it using the Perdew-Burke-Ernzerhof (PBE) exchange-correlation functional, including the D3 correction. An antiferromagnetic structure was considered, where the spins are in the *ab* plane and align in an antiferromagnetic configuration along the *c* axis, as found in our experiment. A 1×1×2 supercell was used for the antiferromagnetic structure. The *k*-point sampling was taken as 8×8×1. Geometries were fully relaxed regarding lattice parameters and interatomic distances until forces were less than 0.001 eV/Å. Using the optimized structure, we utilized the SCAN (strongly constrained and appropriately normed) functional. The Hubbard correction for Cr *d* orbitals was U=3 eV, J=0. The Raman spectrum of $AgCrP_2Se_6$ was calculated using Phonopy and Phonopy Spectroscopy at the PBE+D3 level. The Phonopy package was employed to calculate the zone-center phonon frequencies and phonon eigenvectors of the DFT optimized structures using the finite-displacement approach. The force constant matrix was constructed in a 2×2×1 supercell. The Raman spectrum was simulated by averaging the Raman tensor calculated for each Raman-active phonon eigenmode at the zone center.

Mechanical exfoliation of the $AgCrP_2Se_6$ single crystals was performed using a PDMS stamp. A crystal was first thinned with several Scotch tape exfoliation cycles, followed by thinning onto the PDMS stamp. We then exfoliated the flakes onto cleaned $Si/SiO_2$ (285 nm of $SiO_2$) substrates by placing the PDMS stamp onto the substrate and applying pressure (using ~90 N force) for 20 min followed by slowly peeling the stamp off of the substrate. Several solvent rinses were then carried out to remove as much residue as possible.

Room temperature Raman spectra were collected in two Renishaw inVia Raman microscopes. One of the inVia Raman microscopes was outfitted with a low-frequency module (Coherent/Ondax THz Raman probe) that uses fiber optics to couple 785 nm laser excitation and to direct the scattered light into the inVia spectrometer. The optical layout for this setup is shown in the Supplementary Information Figure S3. The other inVia Raman microscope was used for 633, 514.5 and 488 nm excitation. All spectra were collected by focusing the laser on the exfoliated and bulk crystals through a 50x or 100x objective lenses. All spectra were collected with low excitation powers (<0.1 mW), to minimize laser-induced surface degradation. The CIDs were calculated by baseline subtraction, followed by Lorentzian lineshape fitting. The multi-excitation spectra were collected on thick exfoliated $AgCrP_2Se_6$ flakes on $Si/SiO_2$ substrates, and the spectra were normalized by the Si peak intensities. Reflectance spectra (between 200 – 1100 nm) were collected with a Craic microspectrophotometer. Measurements were performed by placing a freshly cleaved bulk crystal on silicon substrates and by focusing the excitation source (Xe lamp) through a 74x objective lens. The reference used to obtain the Reflectance of the bulk sample is a Ag mirror (Thorlabs), with >95% reflectance between 450 nm to 1100 nm.

**References**

[1] A. N. Rudenko, M. I. Katsnelson, *2D Mater.* **2024**, *11*, 042002.

[2] X. Li, H. Liu, C. Ke, W. Tang, M. Liu, F. Huang, Y. Wu, Z. Wu, J. Kang, *Laser & Photonics Reviews* **2021**, *15*, 2100322.

[3] L. Vannucci, U. Petralanda, A. Rasmussen, T. Olsen, K. S. Thygesen, *Journal of Applied Physics* **2020**, *128*, 105101.

[4] Z. Li, B. Xu, D. Liang, A. Pan, *Research* **2020**, 5464258.

[5] C. Wang, G. Zhang, S. Huang, Y. Xie, H. Yan, *Advanced Optical Materials* **2020**, *8*, 1900996.

[6] X. Zhao, Z. Li, S. Wu, M. Lu, X. Xie, D. Zhan, J. Yan, *Advanced Electronic Materials* **2024**, *10*, 2300610.

[7] A. Nemilentsau, T. Low, G. Hanson, *Physical review letters* **2016**, *116*, 066804.

[8] F. L. Ruta, S. Zhang, Y. Shao, S. L. Moore, S. Acharya, Z. Sun, S. Qiu, J. Geurs, B. S. Y. Kim, M. Fu, D. G. Chica, D. Pashov, X. Xu, D. Xiao, M. Delor, X.-Y. Zhu, A. J. Millis, X. Roy, J. C. Hone, C. R. Dean, M. I. Katsnelson, M. van Schilfgaarde, D. N. Basov, *Nat Commun* **2023**, *14*, 8261.

[9] H. Wang, X. Qian, *Nano letters* **2017**, *17*, 5027.

[10] S. Puebla, R. D'Agosta, G. Sanchez-Santolino, R. Frisenda, C. Munuera, A. Castellanos-Gomez, *npj 2D Mater Appl* **2021**, *5*, 1.

[11] L. Li, W. Han, L. Pi, P. Niu, J. Han, C. Wang, B. Su, H. Li, J. Xiong, Y. Bando, *InfoMat* **2019**, *1*, 54.

[12] X. Ling, S. Huang, E. H. Hasdeo, L. Liang, W. M. Parkin, Y. Tatsumi, A. R. Nugraha, A. A. Puretzky, P. M. Das, B. G. Sumpter, *Nano letters* **2016**, *16*, 2260.

[13] H. B. Ribeiro, M. A. Pimenta, C. J. De Matos, R. L. Moreira, A. S. Rodin, J. D. Zapata, E. A. De Souza, A. H. Castro Neto, *ACS nano* **2015**, *9*, 4270.

[14] G. C. Resende, G. A. S. Ribeiro, O. J. Silveira, J. S. Lemos, J. C. Brant, D. Rhodes, L. Balicas, M. Terrones, M. S. C. Mazzoni, C. Fantini, B. R. Carvalho, M. A. Pimenta, *2D Mater.* **2020**, *8*, 025002.

[15] D. Wolverson, S. Crampin, A. S. Kazemi, A. Ilie, S. J. Bending, *ACS Nano* **2014**, *8*, 11154.

[16] Y. Gong, Y. Zhao, Z. Zhou, D. Li, H. Mao, Q. Bao, Y. Zhang, G. P. Wang, *Advanced Optical Materials* **2022**, *10*, 2200038.

[17] M. Šiškins, M. Lee, F. Alijani, M. R. van Blankenstein, D. Davidovikj, H. S. J. van der Zant, P. G. Steeneken, *ACS Nano* **2019**, *13*, 10845.

[18] Q. Song, X. Pan, H. Wang, K. Zhang, Q. Tan, P. Li, Y. Wan, Y. Wang, X. Xu, M. Lin, *Scientific reports* **2016**, *6*, 1.

[19] A. McCreary, J. R. Simpson, Y. Wang, D. Rhodes, K. Fujisawa, L. Balicas, M. Dubey, V. H. Crespi, M. Terrones, A. R. Hight Walker, *Nano Lett.* **2017**, *17*, 5897.

[20] Q. Song, H. Wang, X. Pan, X. Xu, Y. Wang, Y. Li, F. Song, X. Wan, Y. Ye, L. Dai, *Sci Rep* **2017**, *7*, 1758.

[21] T. Goldstein, S.-Y. Chen, J. Tong, D. Xiao, A. Ramasubramaniam, J. Yan, *Sci Rep* **2016**, *6*, 28024.

[22] S. Zhang, N. Mao, N. Zhang, J. Wu, L. Tong, J. Zhang, *ACS nano* **2017**, *11*, 10366.

[23] S. Zhang, J. Huang, Y. Yu, S. Wang, T. Yang, Z. Zhang, L. Tong, J. Zhang, *Nature Communications* **2022**, *13*, 1254.

[24] M. A. Susner, B. S. Conner, E. Rowe, R. Siebenaller, A. Giordano, M. V. McLeod, C. R. Ebbing, T. J. Bullard, R. Selhorst, T. J. Haugan, J. Jiang, R. Pachter, R. Rao, *J. Phys. Chem. C* **2024**, *128*, 4265.

[25] M. A. Susner, M. Chyasnavichyus, M. A. McGuire, P. Ganesh, P. Maksymovych, *Advanced Materials* **2017**, *29*, 1602852.

[26] E. Bousquet, M. Fava, Z. Romestan, F. Gómez-Ortiz, E. E. McCabe, A. H. Romero, *arXiv preprint arXiv:2406.14684* **2024**.

[27] T. Wang, H. Sun, X. Li, L. Zhang, *Nano Letters* **2024**, *24*, 4311.

[28] R. Pfeiff, R. Kniep, *Zeitschrift für Naturforschung B* **1993**, *48*, 1270.

[29] M. A. Susner, A. Belianinov, A. Borisevich, Q. He, M. Chyasnavichyus, H. Demir, D. S. Sholl, P. Ganesh, D. L. Abernathy, M. A. McGuire, P. Maksymovych, *ACS Nano* **2015**, *9*, 12365.

[30] W. Klingen, R. Ott, H. Hahn, *Zeitschrift für anorganische und allgemeine Chemie* **1973**, *396*, 271.

[31] M. A. Gave, D. Bilc, S. D. Mahanti, J. D. Breshears, M. G. Kanatzidis, *Inorg. Chem.* **2005**, *44*, 5293.

[32] A. S. Pine, G. Dresselhaus, *Phys. Rev. B* **1971**, *4*, 356.

[33] K. Ishito, H. Mao, K. Kobayashi, Y. Kousaka, Y. Togawa, H. Kusunose, J. Kishine, T. Satoh, *Chirality* **2023**, *35*, 338.

[34] K. Ishito, H. Mao, Y. Kousaka, Y. Togawa, S. Iwasaki, T. Zhang, S. Murakami, J. Kishine, T. Satoh, *Nat. Phys.* **2023**, *19*, 35.

[35] S. Streib, *Physical Review B* **2021**, *103*, L100409.

[36] X. Wang, K. Du, W. Liu, P. Hu, X. Lu, W. Xu, C. Kloc, Q. Xiong, *Applied Physics Letters* **2016**, *109*, 123103.

[37] L. Barron, A. Buckingham, *Molecular Physics* **1971**, *20*, 1111.

[38] E. Er, T. H. Chow, L. M. Liz-Marzán, N. A. Kotov, *ACS nano* **2024**, *18*, 12589.

[39] D. Nam, J.-U. Lee, H. Cheong, *Sci Rep* **2015**, *5*, 17113.

[40] R. Roldán, J. A. Silva-Guillén, M. P. López-Sancho, F. Guinea, E. Cappelluti, P. Ordejón, *Annalen der Physik* **2014**, *526*, 347.

[41] W. Luo, A. D. Oyedele, N. Mao, A. Puretzky, K. Xiao, L. Liang, X. Ling, *ACS Phys. Chem Au* **2022**, *2*, 482.

[42] R. Rao, R. Selhorst, R. Siebenaller, A. N. Giordano, B. S. Conner, E. Rowe, M. A. Susner, *Advanced Physics Research* **2024**, *3*, 2300153.

[43] S. Y. Kim, T. Y. Kim, L. J. Sandilands, S. Sinn, M.-C. Lee, J. Son, S. Lee, K.-Y. Choi, W. Kim, B.-G. Park, C. Jeon, H.-D. Kim, C.-H. Park, J.-G. Park, S. J. Moon, T. W. Noh, *Phys. Rev. Lett.* **2018**, *120*, 136402.

[44] B. J. Kip, R. J. Meier, *Appl. Spec.* **1990**, *44*, 707.

[45] J. E. Kim, R. A. Mathies, *The Journal of Physical Chemistry A* **2002**, *106*, 8508.

[46] W. Liu, L. Tang, B. G. Oscar, Y. Wang, C. Chen, C. Fang, *The Journal of Physical Chemistry Letters* **2017**, *8*, 997.

[47] I. Baltog, M. Baibarac, S. Lefrant, *Journal of Physics: Condensed Matter* **2008**, *20*, 275215.

[48] M. A. Susner, M. Chyasnavichyus, A. A. Puretzky, Q. He, B. S. Conner, Y. Ren, D. A. Cullen, P. Ganesh, D. Shin, H. Demir, *ACS nano* **2017**, *11*, 7060.

[49] Z. Yu, R. R. Frontiera, *ACS Nano* **2023**, *17*, 4306.